\documentclass[aps,prl,twocolumn,showpacs,amsmath,amssymb,superscriptaddress]{revtex4}
\usepackage{graphicx} 
\usepackage{bm} 
\renewcommand{\vec}[1]{\bm{#1}} 
 
\begin{document} 
 
\title{Field-induced phase transitions of repulsive spin-1 bosons in optical lattices} 
 
\author{K. Rodr{\'i}guez} 
\affiliation{Institut f\"ur Theoretische Physik, Leibniz Universit\"at Hannover, Appelstrasse 2 D-30167, Hannover, Germany} 
\author{A. Arg{\"u}elles} 
\affiliation{Institut f\"ur Theoretische Physik, Leibniz Universit\"at Hannover, Appelstrasse 2 D-30167, Hannover, Germany} 
\author{A. K. Kolezhuk} 
\affiliation{Institute of Magnetism, National Academy of Sciences and 
  Ministry of Education, 36-b Vernadskii av., 03142 Kiev, Ukraine} 
\affiliation{Institute of High Technologies, T. G. Shevchenko Kiev 
  National University, 64 Volodymyrska str., 01601 Kiev, Ukraine} 
\author{L. Santos} 
\affiliation{Institut f\"ur Theoretische Physik, Leibniz Universit\"at Hannover, Appelstrasse 2 D-30167, Hannover, Germany} 
\author{T. Vekua} 
\affiliation{Institut f\"ur Theoretische Physik, Leibniz Universit\"at Hannover, Appelstrasse 2 D-30167, Hannover, Germany} 
\date{\today} 
 
\begin{abstract} 
We study the phase diagram of repulsively interacting spin-1 bosons 
in optical lattices at unit filling, showing that an externally 
induced quadratic Zeeman effect may lead to a rich physics characterized 
by various phases and phase transitions. We find that the main properties of the system may be described by 
an effective field model, which provides the precise location of the phase boundaries for any dimension, 
being in excellent agreement with our numerical calculations for
one-dimensional (1D) systems.  
Our work provides a quantitative guide for the experimental analysis of various types of 
field-induced quantum phase transitions in spin-$1$ lattice bosons. These transitions, which are precluded in spin-$1/2$ systems,  
may be realized using an externally modified quadratic Zeeman coupling,   
similar to recent experiments with spinor condensates in the continuum. 
\end{abstract} 
 
\maketitle

 


Ultracold atoms in optical lattices constitute a highly controllable 
scenario for the analysis of strongly-correlated systems, as highlighted 
by the realization of bosonic and fermionic 
Mott-insulators~(MIs)~\cite{Greiner2002,Jordens2008,Schneider2008}. 
Interestingly, on-going experiments~\cite{Jordens2010,Molley2010} are approaching the regime at 
which magnetic properties, including the long-pursued Ne\'el antiferromagnet in spin-$1/2$ 
fermions, could be revealed. 
Optically trapped spinor gases, formed by atoms with various Zeeman substates, 
are particularly interesting in this sense. The internal degrees of freedom 
result in an exceedingly rich physics, mostly studied in the context of spinor Bose-Einstein 
condensates~(BECs)~\cite{Ho1998,Ohmi1998,Stenger1998,Barrett2001}. 
Spinor gases in lattices are particularly exciting, since they provide 
unique possibilities for the analysis of quantum magnetism.


Spin-$1$ gases are the simplest spinor system beyond the two-component one. 
Depending on interparticle interactions~\cite{Ho1998,Ohmi1998} (given by 
the $s$-wave scattering lengths $a_{0,2}$ for collisions with total spin $0$ and $2$), 
spin-$1$ BECs present a ferromagnetic~(FM) ground state~(for $a_0>a_2$ as  
in $^{87}$Rb $F=1$~\cite{Barrett2001}) or an antiferromagnetic~(AFM), also called polar, one 
(for $a_2>a_0$, as in $^{23}$Na~\cite{Stenger1998}). 
Spin-$1$ lattice bosons have also attracted a strong interest, especially the AFM case, for which a wealth of quantum phases have been
predicted~\cite{Demler2002,Yip2003,Zhou2003,Imambekov2003,
Imambekov2004,Snoek2004,Rizzi2005,Harada2007,Chung2009}. 
For AFM interactions, in 2D and 3D the MI states at odd filling 
are nematic~\cite{Imambekov2003,Papanicolaou1988,Harada2007}, whereas in 1D 
quantum fluctuations lead to a spontanoeusly dimerized ground 
state~\cite{Demler2002,Yip2003,Zhou2003,Imambekov2003,Rizzi2005,
Chubukov1990,Chubukov1991,Fath1995}. The case $a_0=a_2$  
exhibits an enlarged $SU(3)$ symmetry with a highly degenerate ground state~\cite{Batista2002}. 



\begin{figure}[tb] 
\includegraphics[width=0.42\textwidth]{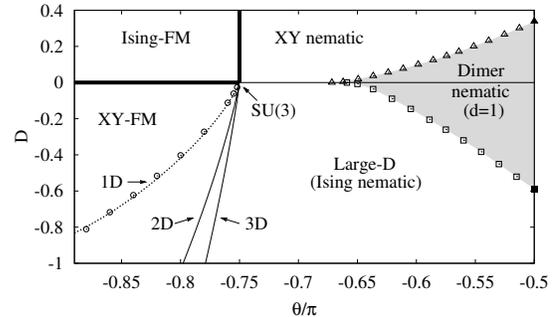} 
\vspace*{-0.5cm}
 \caption{ 
\label{fig:phd} 
Mott phases of spin-$1$ lattice bosons at unit filling, as a function of $\theta\equiv\arctan(J_2/J_1)-\pi$ and 
the QZE $D$. Thick solid lines correspond to first order phase transitions for any $d$.  
 Grey region is the dimerized phase (only in 1D). The 
 symbols represent 1D numerical data (see text). 
 } 
\vspace*{-0.5cm}
\end{figure} 


Most spin-$1$ species are naturally close to 
this $SU(3)$ point (i.e. $a_0\approx a_2$), where small 
external perturbations, as Zeeman shifts, may have a large effect, 
reducing the system symmetry, and thus favoring different phases.
Since interactions preserve the magnetization, ${\cal M}$, the linear Zeeman effect~(LZE) may 
be effectively gauged out~(although the phase diagram depends on ${\cal M}$~\cite{Imambekov2004,Chung2009}). On the contrary, the quadratic Zeeman 
effect~(QZE) plays a crucial role in spinor gases. 
In spite of its importance, the role of the QZE in the quantum phases of 
spin-$1$ lattice bosons remains to a large extent unexplored, with the sole exception 
of the recent 3D mean-field analysis of Ref.~\cite{Chung2009}, where    
it was shown that for finite ${\cal M}$ the QZE
may lead to nematic-to-ferromagnetic~(or partially magnetic) transitions. 


This Letter discusses, for the first time to our knowledge, the complete 
phase diagram (Fig.~\ref{fig:phd}) for MI phases (with unit filling) of spin-$1$ bosons in the presence of QZE, for the experimentally 
relevant case of a balanced mixture, i.e. with ${\cal M}=0$. 
Combining an effective field theory (for any dimension), 1D 
DMRG calculations, and exact Lanczos diagonalization, we obtain the phase boundaries, characterizing 
the phase transitions. 
We note that the QZE may be controlled by means of 
microwave and optical techniques~\cite{Gerbier2006,Santos2007}.
Hence, as recently demonstrated for spinor BECs 
in the continuum~\cite{Sadler2006}, our results show that a controlled 
quenching of the QZE may permit the observation of field-induced phase transitions in 
spin-$1$ lattice bosons, which are precluded by simple use of the LZE due to 
conservation of ${\cal M}$, and thus are absent in 
spin-$1/2$ systems. In addition, optical Feshbach resonances~\cite{Fedichev1996,Papoular2010} 
permit the modification of the ratio $a_2/a_0$, so that the full phase diagram 
discussed below may be explored with state of the art techniques.

 
We consider repulsively-interacting ultracold spin-$1$ bosons in a 
$d$-dimensional hypercubic lattice, prepared in a balanced mixture ($\mathcal{M}=0$). 
The inter-particle interactions are characterized by the coupling constants $g_{0,2}=4\pi\hbar^2a_{0,2}/m_a$~(where $m_a$ is the atomic mass).
At integer filling the system is in the MI regime 
if the (positive) interaction constants $g_{0,2}\gg t$, where $t$ is the hopping amplitude between neighboring 
sites. In second order of perturbation theory in $t$, the low-energy physics is given by super-exchange processes, being 
described by an effective bilinear-biquadratic spin Hamiltonian \cite{Imambekov2003,Yip2003}: 
\begin{equation}  
\label{ham-bilbiq} 
\widehat{H}_I=-\sum_{\langle ij\rangle} \left[ J_{1}\vec{S}_{i}\cdot\vec{S}_{j} 
+J_{2}\big(\vec{S}_{i}\cdot\vec{S}_{j}\big)^{2} \right], 
 \end{equation} 
where $\vec{S}_{i}$ are spin-$1$ operators at lattice site $i$, the sum 
runs over nearest neighbors, $J_{1}=2t^2/g_2$ and 
$J_{2}=2t^2/3g_2+4t^2/3g_0$ (both are positive). 
Note that the FM case~($a_0>a_2$) corresponds to $J_1>J_2$, whereas 
the AFM case~($a_2>a_0$) results in $J_2>J_1$.
As mentioned above, typically $a_0\approx a_2$, 
which corresponds to the vicinity of the ferromagnetic $SU(3)$ point 
($J_{1}=J_{2}$).

As aforementioned, the QZE, characterized by the externally controllable constant $q$, plays a crucial 
role in the physics of the system. The QZE (which resembles the so-called single-ion term in 
other condensed-matter scenarios) leads to a modified Hamiltonian of the form
\begin{equation}  
\label{QZE}  
\widehat{H}= \widehat{H}_I-DJ\sum_{i}\big(S_{i}^{z}\big)^{2}, 
\end{equation} 
where $D=q/J$, with $J\equiv\sqrt{J_1^2+J_2^2}$. 
In the following, we employ $J$ as the energy unit, setting $J=1$, and 
introduce the standard parametrization of the exchange constants  
$J_{1}=-J\cos(\theta) ,\quad J_{2}=-J\sin(\theta)$, where the angle $\theta$ takes values in 
the interval $(-\pi+\arctan\frac{1}{3},-\frac{\pi}{2})$ as the ratio $a_{2}/a_{0}$ varies from $0$ to 
$+\infty$ \cite{Imambekov2003}.


For $\theta<-3\pi/4$ ($J_{1}>J_{2}$) if $D\geq 0$ the ground state is a fully polarized ferromagnet in either $m=\pm 1$ (Ising-FM), and since $\mathcal{M}=0$ 
phase separation into ferromagnetic $m=\pm 1$ domains is expected.
For $D<0$ the ground state for small values of $|D|$ is 
an XY-ferromagnet (XY-FM), i.e. the system fulfills $\langle S_i^z\rangle=0$, but presents a non-zero transversal magnetization. 
This phase is ordered in dimensions $d\ge 2$, exhibiting in 1D a quasi-long-range order 
with leading power-law decay of $xy$ spin correlations. 
For larger $|D|$ (keeping $D<0$) there is a phase transition between the XY-FM and the so-called 
large-$D$ phase (also called Ising-nematic), in which all atoms are in the $m=0$ Zeeman substate, and 
hence all spin correlations decay exponentially. The field-induced phase transition between large-$D$ and XY-FM is discussed in detail below.

For $\theta>-3\pi/4$ ($J_{1}<J_{2}$) the dominant correlations are of spin-nematic (quadrupolar) type~\cite{Papanicolaou1988,Chubukov1991}. 
A XY-nematic phase occurs for $D>0$, characterized for $d\ge 2$ by $\langle (S^{+})^{2}\rangle\neq 0$ and $\langle \mathbf{S}\rangle=0$, 
and in 1D by power-law correlations of the quadrupolar order parameter and exponentially decaying in-plane spin correlations. On the contrary, for $D<0$ 
the large-$D$ phase is favoured. In 1D, for $D=0$ a dimer nematic phase is expected~\cite{Demler2002,Yip2003,Zhou2003,Imambekov2003,Rizzi2005,
Chubukov1991,Fath1995}. 
We show below that in 1D the QZE induces a transition between the dimer-nematic phase and 
the XY-nematic (or large-$D$) one.


To study the phase diagram near the $SU(3)$ point, we develop a low-energy effective 
field theory~\cite{Ivanov2003} based on spin-$1$ coherent states $|\psi\rangle =\sum_{a=x,y,z}(u_{a}+iv_{a})|t_{a}\rangle$ where 
$|t_{a}\rangle$ are three Cartesian spin-$1$ 
states~($|t_z\rangle\equiv|m=0\rangle$, $|m=\pm 1\rangle\equiv \mp(1/\sqrt{2})(|t_x\rangle\pm i|t_y\rangle)$). 
The real vectors $\mathbf{u}$, $\mathbf{v}$ (defined at each lattice site $\mathbf{n}$) satisfy the constraints 
$\mathbf{u}^{2}+\mathbf{v}^{2}=1$, $\mathbf{u}\cdot\mathbf{v}=0$. The vector $\mathbf{u}$ plays the role of 
director vector for the nematic phases discussed below. The average spin on a site fulfills $\mathbf{M}\equiv\langle\psi|\mathbf{S}|\psi\rangle=2(\mathbf{u}\times\mathbf{v})$


The FM region $\theta<-3\pi/4$ is characterized by $\vec{M}$ and the Euler angles 
$(\vartheta,\varphi,\chi)$ which parametrize the orientation of the mutually orthogonal vector pair $(\vec{M},\vec{u})$. 
The Euclidean Lagrangian for the model~(\ref{QZE}) takes then form: 
\begin{eqnarray}  
\label{L-lattice}  
L&=&i\sum_{n}M_{n}\{\partial_{\tau}\chi_{n}+(\partial_{\tau}\varphi_{n})\cos\vartheta_{n}\} 
-D\sum_{n} B_{n}^{zz}\nonumber\\ 
&-&\sum_{\langle 
nn'\rangle} \Big\{ J_{1}\vec{M}_{n}\cdot\vec{M}_{n'}  
+J_{2}\,\text{tr}\big(\mathbf{B}^{T}_{n}\cdot\mathbf{B}_{n'} \big) \Big\}, 
\end{eqnarray} 
where $B^{ab}=\langle \psi|S^{a}S^{b}|\psi\rangle$ is related to the nematic tensor 
$\mathbf{Q}=\frac{2}{3}-\mathbf{B}-\mathbf{B}^{T}$~\cite{DeGennesBook} and the magnetization $M_{a}=\epsilon_{abc}B^{bc}$. 
Assuming $\mathbf{u}$ and $\mathbf{v}$ as smooth fields, we may then pass to the continuum, performing a gradient expansion in 
Eq.~(\ref{L-lattice}). For $D<0$, configurations 
with $\vartheta\simeq \pi/2$  and  $\chi\simeq 0$ are favoured. Since 
$(\cos\vartheta,\varphi)$ and $(M,\chi)$ are pairs of conjugate 
variables,  $\vartheta$ and $\chi$ become ``slaves'' and can be integrated out~(see Ref.~\cite{Ivanov2003} for details). 
The potential energy is 
minimized at 
 $|\mathbf{M}|=M_{0}=\big\{1-\eta^2\big\}^{1/2}$, where $\eta\equiv |D|/2Z(\sin\theta-\cos\theta)$, with 
$Z$ the lattice coordination number. We expand the effective action around 
the equilibrium value, $|\mathbf{M}|=M_{0}+\delta$, integrate out  
$\delta$, and obtain the effective action for $\varphi$, which  can be cast in the 
familiar form of the $(d+1)$-dimensional XY model 
\begin{equation}  
\label{Axy} 
\mathcal{A}_{XY}=(\Lambda^{d-1}/2g)\int d^{d+1}x (\partial_{\mu}\varphi)^{2}.
\end{equation} 
Here $\Lambda$ is the ultraviolet lattice cutoff, and the coupling
constant $g$, acting as an effective temperature, reads
\begin{eqnarray}  
\label{gxy}  
g^{-2}&=&\frac{1}{2Z}\Big\{\frac{1-\eta}{\eta}+\frac{\langle\delta^{2}\rangle}{2\eta^4} \Big\}  
 \Big\{ \Big(1-\eta\Big) \nonumber\\
&\times&\Big( \lambda+1+\eta\Big) 
+\langle 
\delta^{2}\rangle \Big[ 1+\frac{\lambda}{2\eta^{3}}\Big] 
 \Big\}, 
\end{eqnarray} 
where $\langle \delta^{2}\rangle = g_{\delta}C_{d}\pi^{1-d}\int_{0}^{\pi} {dk\,k^{d-1}} 
/{\sqrt{ m_{\delta} ^{2}+k^{2}}}$ is the fluctuation strength, 
$g_{\delta}^{2}=(8Z\eta^{4})/(\lambda+2\eta^2)$ and 
$m_{\delta}^{2}= Z(1-\eta^2)/(\lambda+\eta^2)$, respectively, the coupling constant and mass of the $\delta$ fluctuations, 
$\lambda\equiv \tan\theta/(1-\tan\theta)$, and $C_{d}^{-1}=(4\pi)^{d/2}\Gamma(d/2)$. 
 
Model~\eqref{Axy} describes a phase transition between XY-FM and large-$D$ occuring 
at a non-universal $g=g_{c}$.
For $d=1$ this is a Kosterlitz-Thouless (KT) 
transition, and the XY-FM phase has only a quasi-long-range 
order. For $d\geq 2$, the phase transition belongs to the 
$(d+1)$-dimensional XY universality class, the XY-FM phase is ordered with a spontaneously broken $U(1)$ symmetry ($\varphi=\varphi_0$), and 
the order parameter $\langle \cos\varphi_0 S^x+ \sin\varphi_0 S^y  \rangle\neq 0 $. 
Once $g_c$ is known, Eq.~\eqref{gxy} constitutes an implicit equation 
to determine the transition curve $D(\theta)$. 
The transition line has 
 a \textit{universal slope} for $\lambda\to\infty$ ($SU(3)$ point) given by 
 $\eta=1-O(\lambda^{-1/2})$. 
Fig.~\ref{fig:phd} shows the curve for $d=1$ obtained  
after adjusting the single fitting parameter $g_{c}$ to the numerical results discussed below. An 
excellent agreement with the numerics is obtained for $g_{c}\approx 0.6$.

\begin{figure}[tb] 
 \includegraphics[width=0.48\textwidth]{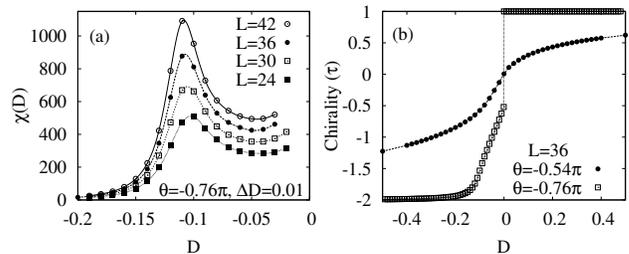} 
 \caption{ 
\label{fig:Susc} 
a) Evolution of the fidelity susceptibility $\chi(D)$ 
at the boundary between XY-FM and large-$D$ phases, with increasing system 
size $L$;  b) Chirality $\tau$ as a function of $D$, for two cuts at the FM ($\theta=-0.76\pi$) and  
 nematic ($\theta=-0.54\pi$) sides. } 
\vspace*{-0.3cm}
\end{figure}

We have numerically evaluated the large-$D$ to XY-FM boundary in 1D by means of 
DMRG calculations~(following the method of Ref.~\cite{Verstraete2004} for up to $42$ sites). 
We found that the most efficient way to locate the phase boundary is to study the fidelity 
susceptibility (FS)~\cite{You2007}, $ \chi(D)=-2\lim_{\Delta D \to 0} \ln\{|\langle 
  \psi(D)|\psi(D+\Delta D)\rangle|^2\} /\Delta D^2$, where the 
  quantity under the logarithm is the 
fidelity, i.e. the Hilbert-space distance between the ground 
states at two values of the QZE coupling, $D$ and $D+\Delta D$.  
Figure\ \ref{fig:Susc}a shows the evolution of the 
peak in $\chi(D)$ with increasing system size. The finite-size 
scaling of the peak position as a function of the number $L$ of sites
follows very accurately a $1/L^2$ law, confirming its KT character. 
Extrapolating the peak position to $L=\infty$ yields the 
curve shown in Fig.~\ref{fig:phd}, which, as mentioned above, agrees perfectly with the 
effective field-theoretical description after fitting the single parameter $g_c$. 


For $d\geq 2$, $g_{c}$ may be estimated by neglecting fluctuations of
$M$ and demanding the critical $|D|$ to match the Ising value 
$ZJ_{1}$ at $J_{2}=0$, which yields $g_c=(8 Z / 5)^{1/2}$. For a square lattice 
the resulting $g_{c}\approx 2.53$  compares favorably with the 
known Monte-Carlo result $g_{c}\approx 2.20$ for the classical 3D XY transition 
on a cubic lattice~\cite{Gottlob1993}. 
The corresponding transition curves obtained from the implicit Eq.~\eqref{gxy} 
are also shown in Fig.~\ref{fig:phd}.
 

For the AFM region, $\theta>-3\pi/4$, the effective theory can be formulated in terms of 
the nematic director $\vec{u}$ \cite{Ivanov2003}, or alternatively via the nematic tensor 
$Q^{ab}=u_{a}u_{b}-\delta_{ab}/3$ 
\begin{equation}  
\label{A-nem-dG}  
 \mathcal{A}_{n}=(\Lambda^{d-1}/4g_{n})\int d^{d+1}x \big\{ 
(\partial_{\mu}Q^{ab})^{2}+2\Lambda^{2}m_{n}^{2}Q^{zz}\big\} 
\end{equation} 
where the coupling  $g_{n}= \sqrt{Z/2\lambda}$ vanishes at the $SU(3)$ point~\cite{Ivanov2003} and  
$m_{n}^{2}=D/\sin\theta$ is the QZE-induced mass. The $\vec{u}$ anisotropy is of the easy-plane (easy-axis) type 
for $D\!>\!0$ ($D\!<\!0$).  For $d\geq2$  there is 
a long range nematic order for any $D$, with a single transition at $D\!=\!0$, between a gapless XY-nematic phase with 
$\langle (S^{+})^{2}\rangle\!\neq\! 0$ at $D\!>\!0$ and a gapped Ising-nematic $\langle (S^z)^{2}\rangle= 0$ at $D\!<\!0$. 

For $d=1$,  at $D=0$ the AFM phase presents exponentially decaying nematic correlations 
(albeit with a very large correlation length $\xi\propto e^{4\pi/g_{n}}$ and 
a tiny spin gap $\Delta\propto \xi^{-1}$ 
close to the $SU(3)$ point~\cite{Chubukov1990,Chubukov1991}). 
In this phase the lowest excitations have total spin $S=2$, which can be understood by noticing 
that in model~(\ref{A-nem-dG}) the magnetization $\vec{M}$ is a composite field, 
$M_{a}\propto\epsilon_{abc}Q^{bd} (\partial_{\tau}Q^{cd})$. 
$Q^{ab}$ can be approximately described as a free 
field with finite mass $m=\Delta^{1/2}$, so its correlator corresponding to   
$S=2$ excitations decays as $x^{-1/2}e^{-mx}$,  
while the spin correlator describing $S=1$ excitations decays 
much faster, as $x^{-2}e^{-2mx}$. 
However, as aforementioned, this featureless disordered-nematic phase \cite{Chubukov1990,Chubukov1991} 
acquires a very weak long-range dimer order all the way to the $SU(3)$ point,
due to the condensation of $\rm Z_{2}$ disclinations~\cite{Zhou2003,Grover2007}.
Although model~\eqref{A-nem-dG} cannot describe dimerization, it may be employed to determine 
the boundaries of XY-Nematic and Ising nematic phases at very small $|D|$, provided that $\rm Z_{2}$ disclinations do not play role at the corresponding phase transitions.
When $D>0$, this transition is KT, and the transition line close to the $SU(3)$ point fulfills 
$D=CJ_{2}\exp\{-4\pi(1-J_{1}/J_{2})^{-1/2} \}$, where $C>0$ is a numerical constant. At $D<0$ the phase transition is Ising-like, 
with a boundary similar as for $D>0$, but a negative constant $\tilde C$. As shown below, this analysis provides a good insight on the dimer-to-nematic transitions.



To characterize numerically the boundaries of the dimerized phase (where FS 
remains featureless), we employed level spectroscopy analysis~\cite{Okamoto1992}. 
In a finite chain, two dimerized ground states 
(degenerate in the thermodynamic limit) split in energy, so the 
lowest excited state in the dimerized phase is unique and belongs to the $\mathcal{M}=0$ sector. In 
contrast, both in the large-D phase~($D<D_{c}^{-}$) and in the 
XY-nematic~($D>D_{c}^{+}$), the lowest excited states are 
twofold degenerate, having $\mathcal{M}=\pm 1$ and $\mathcal{M}=\pm 2$, respectively. Thus, in finite chains a 
level crossing between the lowest excited singlet and doublet states 
occurs when changing $D$.  Our extrapolated results for $D_{c}^{\pm}$, 
obtained by Lanczos diagonalization for periodic systems of up to 
$L=16$ sites, are shown in Fig.~\ref{fig:phd} with $\triangle\to D_c^+, \Box\to D_c^-$. Note that when approaching the $SU(3)$ point 
our numerics cannot recover the exponentially small dimer region, which basically 
reduces to $D=0$ line. The finite size 
extrapolation of $D_{c}^{+}$ follows $1/L^2$ law, confirming its 
KT nature.
 
To determine the universality class of the $D=D_{c}^{-}$ transition, we have computed the 
central charge at $D=0$, $\theta=-0.73\pi$~\cite{footnote-1}. 
The block entanglement entropy for an open 1D system of size $L$, divided into 
two pieces of size $l$ (block) and $L-l$ (environment), behaves as 
$\mathcal{S}=\frac{c}{6}\log\left[\frac{L}{\pi }\sin\left(\frac{\pi 
    l}{L}\right)\right]+A$,  
where $c$ is the central charge and $A$ is 
a non-universal constant~\cite{Korepin2004,Calabrese2004}. Setting $l=\frac{L}{2}$, following 
Ref.~\cite{Tagliacozzo2008}, and using DMRG to evaluate $\mathcal{S}$ for several $L$ values, 
we obtain $c\simeq1.5$. The $D=D_{c}^{+}$ KT line 
 has $c=1$; subtracting its contribution, we get 
$c=\frac{1}{2}$ for the $D=D_{c}^{-}$ line, confirming its Ising nature.

 
Finally, we discuss the behavior of the chirality $\tau=\frac{1}{L} 
\sum_i (\hat{n}_{i,+1}+\hat{n}_{i,-1}-2\hat{n}_{i,0})=\frac{1}{L}\sum_{i}\big\{3(S_{i}^{z})^{2}-2\big\}$, which 
can be easily monitored in Stern-Gerlach-like time-of-flight 
experiments. Figure~\ref{fig:Susc}b shows $\tau$  
as a function of $D$. 
At the FM side, $\tau$ is discontinuous at $D=0$ 
indicating the first-order character of the transition, which is clear since  
for the Ising-FM phase~($D>0$) $\tau=1$, while in the XY-FM 
phase ($D<0$) for $D\to -0$ the ground state energy is minimized at 
$(M,\vartheta,\chi)=(1,\frac{\pi}{2},0)$, and thus for $\langle 
(S^{z})^{2}\rangle\to \frac{1}{2}$ and $\tau\to -\frac{1}{2}$. 
At the XY-FM to large-D transition, $\tau$ 
practically saturates to $-2$ for a value of $D$ close to that obtained from 
the FS analysis. On the AFM side the limit $D\to 0$ is non-singular and 
thus $\tau=0$ there. The nematic-to-dimer transitions do not present any pronounced 
feature of $\tau$. These transitions could be revealed experimentally 
by Faraday rotation techniques~\cite{Eckert2008} or those recently explored 
in Ref.~\cite{Trotzky2010}.

In summary, we have obtained the complete phase diagram (for any dimension) for spin-$1$ lattice bosons 
in the MI phase (at unit filling) in the presence of quadratic Zeeman coupling. Our results provide hence 
a quantitative guide for the analysis of field-induced quantum phase transitions in lattice bosons, which, 
similar to recent experiments with spinor BECs in the continuum~\cite{Sadler2006}, may be realized modifying the 
QZE by means of microwave dressing. Starting in the large-D phase, and dynamically 
modiying the QZE across the transitions discussed in this paper, should result in the FM regime 
in the appearance of XY-FM domains, similar to those observed in spin-$1$ BECs~\cite{Sadler2006}, whereas 
quenches in the AFM regime should lead to nematic domains with different 
$\langle (S_{x,y})^2\rangle$ but homogeneous $\langle \mathbf{S}\rangle=0$. We stress that 
such field-induced transitions are precluded for spin-$1/2$, constituting an interesting novel feature of lattice spinor gases.

\acknowledgments
We thank A. Honecker for useful discussions. 
A.K. acknowledges the hospitality of the 
Institute of Theoretical Physics at the Leibniz University of Hannover, and support by Grant 220-10 from the Ukrainian Academy of 
Sciences. This work has been supported by the Center for Quantum 
Engineering and Space-Time Research (QUEST), and the SCOPES Grant IZ73Z0-128058.



\begin{thebibliography}{10.}
\bibitem{footnote-1} Strictly speaking, this point is inside the dimerized phase, but since the correlation 
length is extremely large, the system can be considered as gapless.
\end{thebibliography}
\end{document}